# The Ultraviolet View of Star- and Planet-Formation: Disks, Accretion, and Outflows with the Hubble Space Telescope into the 2030s

Kevin France[1], Eric Gaidos[2] , Catherine Espaillat[3], Carlo F. Manara[4], Edwin Bergin[5]
**Endorsers:** Fatemeh Zahra Majidi[6], Hans M. Günther[7], J. Serena Kim[8], Juan Manuel Alcalá[6], Connor Robinson[9], Amelia Bayo[4], Matthew Kalscheur[1], Frederick M Walter[10], Miguel Vioque[4], Caeley V. Pittman[3], Deirdre Coffey[11], C.M. Johns-Krull[12], Jesús Hernández[13] , Gabriella Zsidi[2], Aurora Sicilia-Aguilar[14], P.C. Schneider[15], Hsien Shang[16], Joel Kastner[17], Nicole Arulanantham[18], Nicolas Grosso[19]

**Affiliations:** (1) LASP & Dept. of Astrophysical and Planetary Sciences, CU Boulder, USA (2) Dept. of Earth Sciences, U. Hawai'i, USA; (3) IAR & Dept. of Astronomy, Boston U., USA; (4) European Southern Observatory, Germany; (5) Dept. of Astronomy, U. Michigan, USA; (6) INAF-OACN, Italy; (7) Kavli Institute for Astrophysics & Space Research, MIT, USA; (8) Steward Observatory, U. Arizona, USA; (9) Dept. of Physics & Astronomy, Alfred University, USA; (10) Dept of Physics & Astronomy, Stony Brook University, USA; (11) School of Physics, University College Dublin, Ireland; (12) Dept. of Physics & Astronomy, Rice University, USA; (13) Instituto de Astronomía, UNAM, México; (14) SUPA, School of Science and Engineering, University of Dundee, UK; (15) Institute of Theoretical Physics and Astrophysics, Kiel University, Germany; (16) Institute of Astronomy and Astrophysics, Academia Sinica, Taiwan; (17) Lab for Multiwavelength Astrophysics, Rochester Institute of Technology, Rochester, USA; (18) Astrophysics & Space Center, Schmidt Sciences, USA; (19) Aix-Marseille Univ, CNRS, CNES, LAM, Marseille, France

**Executive Summary:** The spatial distribution and lifetime of molecular gas in the inner regions of young circumstellar disks are key to understanding the formation of planetary systems. Gas-rich disks are observed to disperse in the first ~10 Myr, and recent observational and theoretical evidence suggests that circumstellar disks winds may dominate the removal of angular momentum from the disk, allowing it to dissipate through accretion onto the central star and through low-velocity (<~ 30 km/s) outflows. The Hubble Space Telescope has revolutionized our understanding of the disks, accretion, and outflow processes that drive the evolution of planet-forming disks and is poised to answer the key questions in the field in the coming decade. We describe how HST's ultraviolet capabilities can address these questions and identify key goals and high-priority observations for HST into the 2030s.

## 1. Key Science Questions Requiring Hubble's UV Capabilities

The protoplanetary disks around the youngest ($\lesssim$ 10 Myr) stars consist of both gas and dust. Although the overall mass of gas is $\sim$100 times that of the dust, readily-detected infrared (IR) emission from dust in the inner disk ($\lesssim$ 1 au) has made it a widely studied aspect of these systems. Cold gas in outer disks is observed at mm wavelengths by ALMA (e.g., Miotello et al. 2023), and hot, shocked gas in accretion flows or on the star is studied by its profuse atomic line and continuum emission (e.g., Pittman et al. 2022, Thanathibodee et al. 2019). Investigation of warm molecular gas in the inner disk has lagged that of dust, but CO and $H_2O$ have been studied by 2–5 μm spectroscopy from the ground (Banzatti et al. 2017). Space telescopes enable more sensitive studies of other molecules in the mid-IR, first with Spitzer (Pontoppidan et al. 2010), and now JWST (Arulanantham et al. 2025 and references therein).

Ultraviolet spectroscopy is a powerful and complementary tool for observing molecular gas in inner protoplanetary disks. There are strong electronic band systems of $H_2$ and CO in the 110–170 nm bandpass (Fig. 1, Herczeg et al. 2002; France et al. 2011a). UV spectroscopy has also been crucial in studies of the energetic exchange of mass and momentum between protostars and their disks, driving disk evolution and affecting planet formation (Hartmann et al. 2016; Pascucci et al. 2023, Manara et al. 2023). Data at $\lambda$<300 nm directly assay excess emission from accretion (Calvet & Gullbring 1998; Pittman et al. 2022, 2025), and can trace outflows (Xu et al. 2021). The ULLYSES survey (Roman-Duval et al. 2025) demonstrated the power of HST for UV spectroscopy for such science (see also Espaillat et al. 2022), motivating studies of young stellar objects over a wider range of stellar masses, ages, and disk evolutionary states.

The unique UV capabilities of HST can continue to contribute into the 2030s to resolving outstanding questions in star and planet formation research, including:

**Q1: What are the mass outflow rates of protoplanetary disk winds and what mechanisms drive them?** Winds are central to a paradigm shift in our understanding of how mass and angular momentum are transported through protoplanetary disks, and could drive accretion where ionization levels are insufficient for magneto-rotational instability to operate (Pascucci et al. 2023). These winds can arise from multiple locations in the disk, and observations of multiple spectral indicators with sufficient velocity resolution are needed to resolve this structure (Xu et al. 2021; Arulanantham et al. 2024).

**Q2: What are metal abundances in accretion flows and do these reflect selective sequestration of elements in dust trapping or planetesimals?** Modeling of key atomic emission lines in UV spectra can constrain the abundances of major metals and any depletion relative to the standard abundances for nearby star-forming regions (McClure et al. 2020, Micolta et al. 2023, Thanithibodee et al. 2024, Micolta et al. 2024). Connecting depletion with structures (as potential traps or signposts of planet formation) in the outer disks or its evolutionary state could provide important insight into the process of planet formation. Far ultraviolet lines of highly ionized volatile elements (i.e. C and N) derive primarily from accretion flows (as opposed to stellar active regions), and comparison of their depletion relative to that of refractories such as Si and Mg could constrain the temperature at which sequestration occurs.

**Q3: What is the distribution of lowest accretion rates among young stellar objects?** In the classical (idealized) picture of viscous disk accretion, accretion rates decline smoothly as an inverse power-law with time (e.g., Armitage 2011). However, departures from this secular evolution are expected because of MHD winds, photoevaporative mass loss, and changes in disk processes (e.g., Gaidos et al. 2025) and there is evidence for a lower cut-off at $\sim 10^{-10}$ $M_{sun}$ $yr^{-1}$ (Thanathibodee et al. 2023). Sensitivity to low accretion rates also requires accurate subtraction of a proxy star's UV emission, and hence spectroscopy.

**Q4: What is the prevalence of molecular hydrogen gas in the inner regions of transition and non-accreting disks, and very young debris disks?** A residual gas disk during the late stages of rocky planet formation can be accreted and dissolved into magma oceans, influencing planet interior chemistry (Rogers et al. 2024) and secondary atmospheres. These extended primordial atmospheres can also amplify accretion of residual planetesimals (Takayuki & Kejii 2010). Spectra of UV-fluorescent $H_2$ ("UV-H2") are sensitive to gas surface densities lower than $10^{-6}$ g $cm^{-2}$, making them an extremely useful probe of remnant gas at $r < 10$ AU during the disk dispersal stage when mid-IR CO spectra or traditional accretion diagnostics (e.g., H$\alpha$ equivalent widths) may not avail (Ingleby et al. 2011; France et al. 2012b; Arulanantham et al. 2018; Alcala et al. 2019, Skinner & Audard 2022).

**Q5: How do accretion flows change with time?** T Tauri accretion is highly variable, but the mechanism(s) driving this variability and the back-reaction of variability accretion-powered high-energy radiation on the disk are largely unknown and active areas of investigation (Fischer et al. 2023). Significant changes in accretion luminosity will move the water ice line, potentially affecting formation and volatile incorporation of planets (Houge & Krijt 2023, Ros & Johansen 2024). Different regions (funnel flow, shock, and post-shock gas) are probed by different wavelengths/lines and their variability is expected to be asynchronous, Accretion during a burst can be mapped using line profiles vs temperature:

ULLYSES provided the most robust FUV/NUV study to date (Wendeborn et al. 2024), but was limited to only four stars. Similar observations of more stars with a range of magnetospheric structures are needed.

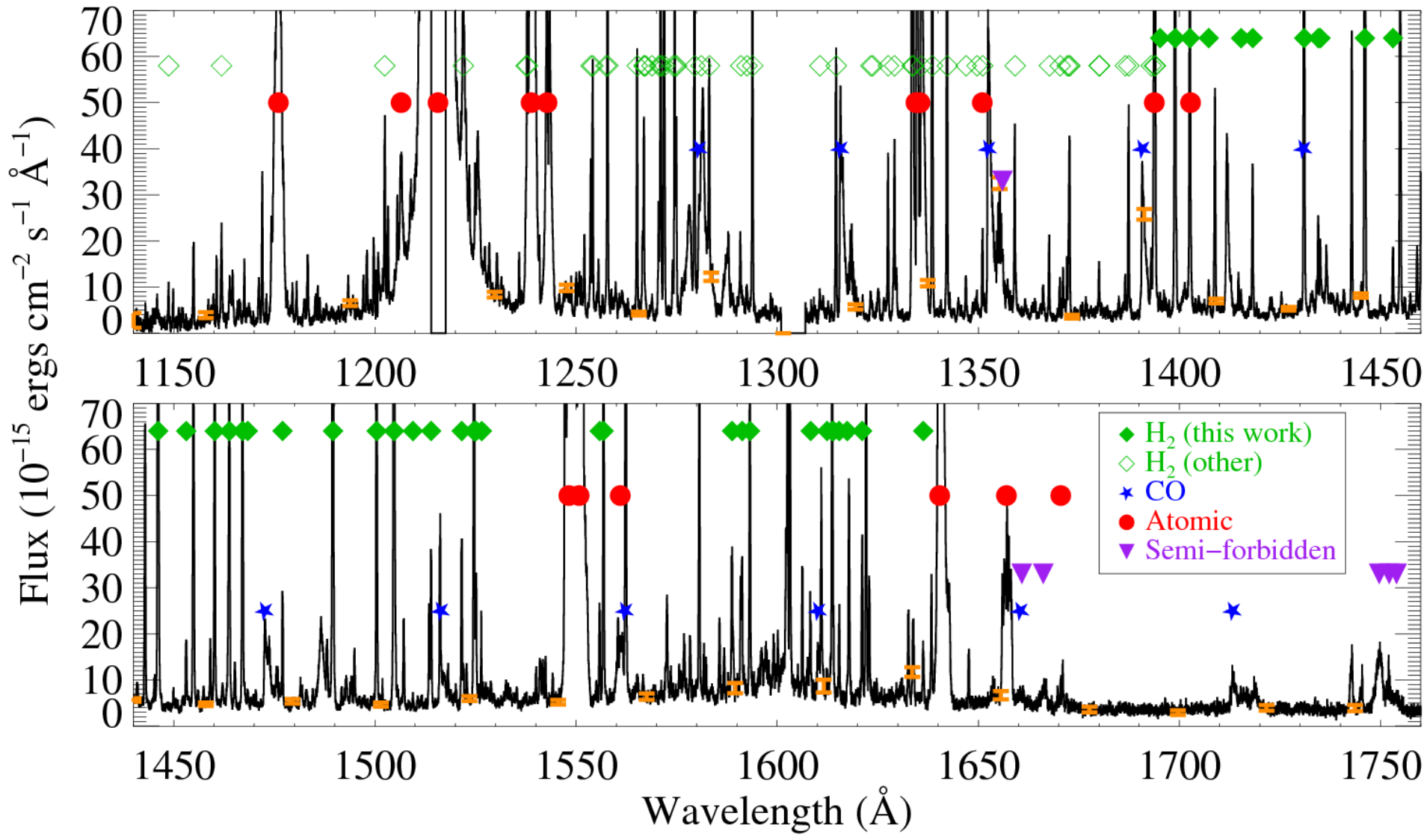

*Figure 1: Representative far-ultraviolet molecular spectrum of a protoplanetary disk, with $H_2$ and CO fluorescent emission lines noted in green and blue, and strong resonance and semi-forbidden lines noted in red and purple. Figure adapted from France et al. (2023).*

**2. Instrument and Operational Requirements**

**Disk winds:** Molecular disk wind kinematics are best studied with **HST-COS M-mode FUV spectroscopy**. Line profiles show that UV-$H_2$ emission from T Tauri stars originates at $0.1 <\sim r \lesssim 10$ au in the disk (Herczeg et al. 2004; France et al. 2012a; Hoadley et al. 2015) and slow molecular winds (Kalscheur et al. 2025; Fig. 2). Kinematic information complements spatial morphologies from JWST (e.g., Fig, 2), particularly for targets with $H_2$ data cubes (see Section 4). Molecular disk winds can also be spectrally imaged using **HST-STIS M- and L-mode imaging FUV spectroscopy (G140M, G140L)** and $H_2$ maps generated using **HST-ACS/SBC FUV filter imaging (F125LP, F150LP, F165LP)**.

**Metals and depletion in accretion flows:** Sufficient spectral resolution of FUV lines is critical to adequately deconvolve the narrow (few tens of km/s) and broad (100s of km/s) components of emission from accretion flow; the former arises from the post-shock region of the flow, is optically thin and is relatable to elemental abundances, while the latter is optically thick (Thanathibodee et al. 2024). Relevant lines include the C IV (154.8 and 155 nm), N V (123.9 and 124.3 nm), and Si IV (139.4 and 140.3 nm) doublet, thus **HST-COS M-mode FUV spectroscopy** is needed.

**The limits of protostellar mass accretion:** Combined NUV - blue optical spectroscopy provides the most robust estimates of the mass accretion rate, including simultaneous constraints on the effects of interstellar and circumstellar reddening (e.g., Ingleby et al. 2013), therefore, **HST-STIS L-mode NUV and optical spectroscopy (G230L, G430L)** are crucial.

**Primordial and second-generation gas in highly evolved and debris disks:** The FUV transitions of $H_2$ are intrinsically strong ($A_{ij} \gtrsim 10^8$ s$^{-1}$) and are photo-excited by Lyα photons (Brown et al. 1981; Herczeg et al. 2002; France et al. 2012b), the strongest stellar emission line in the FUV (Schindhelm et al. 2012). $H_2$ emission lines are seen in every actively accreting YSO observed (see Figure 1 France et al. 2012b, 2023). **HST-COS M-mode FUV spectroscopy** can detect trace amounts of $H_2$ and reconstruct the Lyman-alpha pumping emission (Ingleby et al. 2011; Alcala et al. 2019).

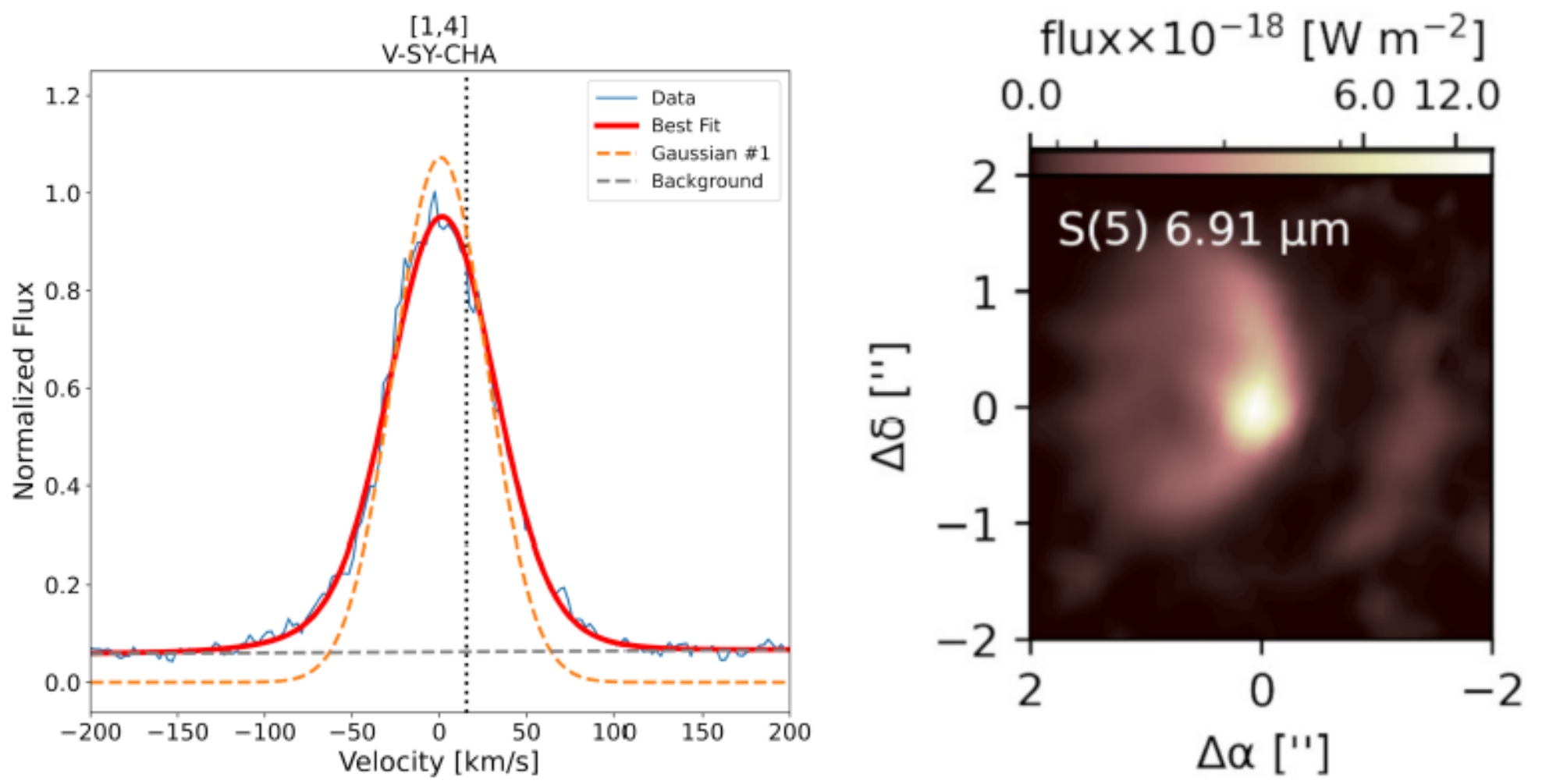

*Figure 2: (Left) Co-added line profile of UV $H_2$ emission from the moderately-inclined disk SY Cha disk, demonstrating a blueshift of -15 km $s^{-1}$ relative to the stellar radial velocity (black dotted line) (Kalscheur et al. 2025). (Right) JWST spectral image of SY Cha in a rotational line of $H_2$ (Schwartz et al. 2025).*

**Mapping accretion flows with variability:** The accretion flow and shock is seen in cool and hot lines. Reverberation mapping between Ly-alpha and $H_2$ emission would allow us to pinpoint the location of the warm $H_2$ gas in the disk. The optimal program would use **HST-COS G130M observations** (Ly-alpha, $H_2$, C II, C III, Si IV, N V) and **HST-STIS L-mode NUV and optical spectroscopy (G230L, G430L)** of several sources in the continuous viewing zone to monitor a star constantly for ~10 days (160 orbits), sufficiently long to capture and cover these 1–2-day bursts.

## 3. Synergies with Current and Future NASA Observatories

**JWST:** The relative contributions of magneto-hydrodynamic (MHD) and photoevaporative (PE) winds, wind kinematics, and the spatial extent of fluorescent $H_2$ winds remain poorly constrained (Kalscheur et al. 2025; Pascucci et al. 2023; Nakatani et al. 2026). While nested jet and wind morphologies and semi-opening angles in JWST imaging are consistent with an MHD origin in some disks (Pascucci et al. 2025), hydrodynamical modelling suggests that PE winds are able to recreate the radial extents, semi-opening angles and line fluxes of observed IR-H2 winds (Nakatani et al. 2026). Comprehensive mapping of UV-H2 emission around a sample of T Tauri stars will help break the impasse. Creating $H_2$ maps with HST-STIS G140M 52″ x 0.2″ long-slit aperture, with velocity centroids on the order of 3–5 km $s^{-1}$ , would provide key kinematic information for disk wind morphology studies.

**UVEX:** The Ultraviolet Explorer (UVEX), slated to be launched in 2030, will conduct a 500-day all-sky synoptic survey in FUV (139-190 nm) and NUV (203-270 nm) bands that is 100 times more sensitive than GALEX (Kulkarni et al. 2021, Fucik 2024). It will identify more and potentially more diverse YSOs, as well as variable YSOs, e.g., due to accretion variability (Zsidi et al. 2025) or dynamically driven accretion outbursts like DQ Tau (Fiorellino et al. 2022). UVEX spectra will be of insufficient resolution and sensitivity for the science presented here, but could motivate follow-up with HST.

**Habitable Worlds Observatory (HWO):** Mapping of UV-$H_2$ molecular disk winds is part of the Compelling Science Case package for HWO (HWO CSIT, priv. comm.). HST spectral maps can serve as a demonstration dataset for HWO analysis pipelines, and importantly, constrain the spatial extent, brightness, and velocity distribution of UV-$H_2$ outflows around young stars. These observables are the precursor science products that will inform the sensitivity, field-of-view and spectral resolution goals for HWO's integral field and multi-object spectrographs as the instrument suite is refined at the end of this decade.

## 4. Spectral and Spatial Mapping of $H_2$ Disk Winds as a Representative Major Initiative

Improved velocity information relative to JWST will allow for the use of wind velocity as a diagnostic of MHD or PE winds. New spatial information on the UV-H2 gas will allow us to determine if IR- and UV-$H_2$ are parts of the same outflow and measure the kinematics of the molecular outflows. Due to the brightness of the typical T Tauri star target and the goal of creating spectral data cubes of the disk wind using pushbroom mapping, observing times for bright disks are of order 10 orbits and 20-30 orbits for typical disks. This means that to develop a statistical sample of UV molecular disk wind maps (e.g., ~half of the ULLYSES sample or most of the UV-bright disks with JWST near- and mid-IR spectral maps), investments of hundreds of orbits are required.

## 5. Community-Wide Recommendations

**Promoting studies of YSO variability:** T Tauri stars are highly variable on timescales of minutes to years due to flaring, evolution and rotation of active regions on the central star, changes in disk accretion rate (Zsidi et al. 2022), and shadowing by structures near the inner disk edge (e.g., Espaillat et al. 2011; Benisty et al. 2023). This variability could impact disk and protoplanet evolution, and serve as a probe of disk structure and chemistry (e.g., Smith et al. 2025). To support UV time-domain investigations, STScI should encourage, identify, and facilitate competitive selection of large (~50 orbit), multi-visit campaigns on individual systems.

**Reassessing limits on target selection for UV spectroscopy:** Well-justified restrictions on observations of UV-bright or potentially UV-variable stars (e.g., Instrument Science Report COS 2017-01) have conserved HST's instruments for decades of impactful science, but have led to gaps in our knowledge of (flaring) young stars and M dwarfs. Substantial improvement in our understanding of flare statistics and age and mass dependence brought about by Kepler and TESS (e.g., Mamonova et al. 2025) as well as consideration that the likely horizon for operation is half a decade, merits a technical reassessment of these limitations.

**Robust funding for HST GO and AR programs:** A broad segment of the astronomical community studying UV astronomy rely on HST program support of graduate students and postdocs. The HST experience continues to shape future leaders in the field, and it is essential that this key role of one of our great observatories be maintained.